\documentclass[twocolumn,preprintnumbers,amsmath,amssymb]{revtex4}

\usepackage{graphicx}
\usepackage{dcolumn}
\usepackage{bm}

\begin{document}

\preprint{OUTP 05 12P}

\title{On the $\Delta I = 1/2$ Rule in Holographic QCD}

\author{Thomas Hambye$^{a,b}$, Babiker Hassanain$^a$, John March-Russell$^a$, and Martin Schvellinger$^a$}
\affiliation{%
$^a$Rudolf Peierls Centre for Theoretical Physics, Department of Physics, University of Oxford, 1 Keble Road,
Oxford, OX1 3NP, UK \\
$^b$IFT-UAM/CSIC, Fac. Ciencias, Universidad Aut\'onoma de Madrid, Cantoblanco, 28049 Madrid, Spain}

\begin{abstract}

We study the $\Delta I = 1/2$ rule for kaon decays
and the $B_K$ parameter for $K^0 - {\bar K}^0$ mixing 
in a dual 5-dimensional holographic QCD model.
We perform, in the chiral limit, computations of the
relevant four-point current-current
correlators, which depend upon self-interactions among the 5D bulk fields.
Spontaneous chiral symmetry breaking ($\chi$SB) is realized through boundary
conditions on the bulk fields.  Numerical results are analyzed
in comparison with QCD, chiral perturbation theory ($\chi$PT)
and data, finding reasonable agreement 
with the experimental values of the $g_8$ and $g_{27}$ parameters
describing the $\Delta I =1/2,3/2$ decay channels.

\end{abstract}

\maketitle

\section{Introduction}

The gauge/gravity duality \cite{Maldacena:1997re,Gubser:1998bc}
provides new tools to investigate non-perturbative QCD. 
Holographic dual models which
allow the study of confinement and $\chi$SB have been
developed in essentially two different approaches: models based 
on string theory/supergravity in 10D \cite{Kruczenski:2003be}, or 
phenomenologically inspired holographic dual models in 5D
\cite{Erlich:2005qh,DaRold:2005zs,Hirn:2005nr,Ghoroku:2005vt,Katz:2005ir}.  
In the past year, a study of meson masses and decay constants in the 5D
models have shown quite good agreement with data, at least for 
the leading meson excitations.  These encouraging
results are based on two-point current-current correlators, and in the
5D description the leading contributions to such correlators
are obtained from bare 5D propagators.  One can also compute
$n$-point correlators at
tree level; however, the fact that 5D interactions must be taken
into account makes an important difference that we shall
investigate here.  
The holographic calculation of four-point
correlators is a non-trivial test for 5D dual models of
non-perturbative QCD, with, perhaps, the most interesting aspect
being the holographic computation of observables which are
important in QCD, particularly the celebrated $\Delta I=1/2$ rule for 
kaon decays.
Since our calculation is the first within this framework,
we focus upon the simplest AdS/QCD model.

\section{The $\Delta I=1/2$ rule, and the $B_K$ mixing parameter}

We start with a review of the relevant facts.
Neglecting CP-violation there are two independent $K^0$ decays: 
$K^0 \rightarrow \pi^+ \pi^-$ and $K^0 \rightarrow \pi^0 \pi^0$. 
These two decays are combinations 
of $\Delta I =1/2$ and $\Delta I = 3/2$ isospin amplitudes
($A_0$ and $A_2$, respectively).  
Experimentally $Re A_0/ Re A_2=22.2$, and the unexpected largeness of this
ratio is the $\Delta I = 1/2$ rule. In the chiral limit, and 
at ${\cal{O}}(p^2)$ in the chiral expansion, these two amplitudes 
are expressed in terms of the $g_8$ and $g_{27}$ parameters (see 
e.g.~\cite{KMW,Hambye:2003cy}).  In the following we will also 
discuss the related but simpler observable, $\hat{B}_K$, parameterizing 
$K^0-\bar{K}^0$ mixing.  Performing an OPE, its calculation reduces to that of 
the hadronic matrix element $<\bar{K}^0|Q_{\Delta S=2}|K^0>(\mu)\equiv 
\frac{4}{3}f_{K}^2 M_{K}^2 B_{K}(\mu)$ of the four-quark 
left handed current 
operator $Q_{\Delta S=2}\equiv 4 \eta_{\mu\nu}L^\mu_{{\bar s} d}L_{{\bar s} d}^\nu$ 
with $L^{\mu}_{{\bar q}_1 q_2} 
\equiv {\bar q}_1 \gamma^\mu \frac{(1-\gamma_5)}{2} q_2 $. 
Similarly, $g_{8,27}$ involve
the $K \rightarrow \pi\pi$ matrix elements of 
two $\Delta S =1$ 
operators four-quark left handed current 
operators (above the charm threshold), 
$Q_1\equiv 4 \eta_{\mu\nu}L^\mu_{{\bar s} d}L_{{\bar u} u}^\nu$ and 
$Q_2\equiv 4 \eta_{\mu\nu}L^\mu_{{\bar s} u}L_{{\bar u} d}^\nu$. 

The factorized contribution for these matrix elements is well known; the
unfactorized contribution is more controversial.  In the 
chiral limit the leading $N_c$ unfactorized contributions reduce to the 
calculation of integrals in $Q^2$ (with $Q$ the Euclidean 
momentum flowing between the left handed currents) of two distinct 
four-point current
correlators \cite{PR,Hambye:2003cy}. For $\hat{B}_K \equiv B_K (\mu)
\cdot C_{\Delta S=2}(\mu)$,
\begin{eqnarray}
&&B_K(\mu)=
\frac{3}{4}-\frac{3}{4}\frac{1}{32\pi^2 F_\pi^2} \int
dQ^2 \, W_{LRLR}(Q^2)(\mu), \label{BKint}\\
&&{W}_{LRLR}(Q^2)=
 - \frac{Q^2}{3 F_\pi^2}\eta_{\alpha\beta}\eta_{\mu\nu} 
\int \frac{d\Omega_{q}}{4 \pi} {W}_{LRLR}^{\mu\alpha\nu\beta}(q), \nonumber \\
&& {W}_{LRLR}^{\mu\alpha\nu\beta}(q)=\lim_{l\rightarrow 0}\ i^3 \int d^4x d^4y
d^4z\ e^{iqx+il(y-z)} \nonumber \\
&&~~~~\langle 0\vert T\{
L_{\bar{s}d}^{\mu}(x)R_{\bar{d}s}^{\alpha}(y)
L_{\bar{s}d}^{\nu}(0)R_{\bar{d}s}^{\beta}(z)\}
\vert 0\rangle\vert_{\mbox{\rm\tiny conn}} , \label{WLRLR}
\end{eqnarray}
while $g_8$ and $g_{27}$ are given by 
\begin{eqnarray}
&&g_8 (\mu) = z_1(\mu) \biggl(\frac{4B_K(\mu)}{5} -1 \biggr) + \\
&&~~~~z_2(\mu)\biggl(1-\frac{8B_K(\mu)}{15} - 
\int dQ^2 \frac{{W}_{LLRR}(Q^2)}{4 \pi^2 F_\pi^2}(\mu) \biggr), \nonumber \\
&&g_{27}(\mu)=\biggl(z_1(\mu)+z_2(\mu)\biggr) \frac{4B_K(\mu)}{5}. 
\label{int2}
\end{eqnarray}
In the above, $C_{\Delta S=2}(\mu)$ and $z_{1,2}(\mu)$ are the 
Wilson coefficients
of $Q_{\Delta S=2}$ and $Q_{1,2}$,
$R^{\mu}_{{\bar q}_1 q_2} \equiv 
{\bar q}_1 \gamma^\mu \frac{(1+\gamma_5)}{2} q_2$,
and  
\begin{eqnarray}
&&{W}_{LLRR}(Q^2)=-\frac{Q^2}{3 F_\pi^2}
\eta_{\alpha \beta} \eta_{\mu\nu}
\int \frac{d\Omega_{q}}{4 \pi} {W}_{LLRR}^{\,\mu\,\nu\,\alpha\,\beta}(q), \nonumber \\
&&{W}_{LLRR}^{\,\mu\,\nu\,\alpha\,\beta}(q) = \lim_{k\rightarrow 0}
\ i^3\int d^4x d^4y d^4z\ e^{iqx + ik(y-z)} \nonumber \\
&&~~~~\langle 0\vert T\{L_{\bar{s}u}^{\mu}(x)L_{\bar{u}d}^{\nu}(0)
R_{\bar{d}u}^{\alpha}(y)R_{\bar{u}s}^{\beta}(z)\}\vert
0\rangle\vert_{\mbox{\rm\tiny conn}} .
\label{LLRRint}
\end{eqnarray}

\section{The holographic QCD model}

To compute the four-point correlators through the gauge/gravity
correspondence, we consider the $L^{\mu}_{{\bar q}_1 q_2}$
and $R^{\mu}_{{\bar q}_1 q_2}$ 4D quark currents.
By virtue of the AdS/CFT correspondence, these operators couple to the boundary
values of 5D gauge fields $L^\mu(x^\nu, z)$ and $R^\mu(x^\nu, z)$, and 
the 5D gauge fields are massless.
The 5D action is a $SU(3)_L \times SU(3)_R$ 
Yang-Mills theory 
\begin{equation}
-\frac{M_5}{4} \int \mathrm{d}z \, \mathrm{d}^4x \, \sqrt{g} \,
 (L^a_{MN} L^{a, MN} + R^a_{MN} R^{a, MN}), \label{action}
\end{equation}
where $M_5$ is an undetermined mass scale, and $M=(\mu, 5)$. The fifth
coordinate $z$ runs from $L_0$ to $L_1$,
which are the positions of the UV and IR branes respectively. 
It is possible to set $L_0=0$ as in Refs.\cite{Erlich:2005qh,DaRold:2005zs}
as our results are smooth as $L_0\rightarrow 0$. 
We use the AdS$_5$ metric 
$ds^2 = a(z)^2 (\eta_{\mu\nu} \, dx^\mu dx^\nu - dz^2)$,
where $a(z)^2=(L/z)^2$ and $\eta_{\mu\nu}$ is given by $(+,-,-,-)$. The field strengths are 
$L_{MN}=\partial_ML_N-\partial_NL_M-i [L_M,L_N]$, 
where $L_M=L_M^a T^a$, and similarly for $R_{MN}$. We normalise
${\rm tr}(T^a T^b)=\delta^{ab}$, and define the vector and axial
gauge bosons as $V_M = (L_M + R_M)/\sqrt{2}$ and $A_M = (L_M - R_M)/\sqrt{2}$, respectively.

The holographic dual of this 5D theory is a
4D theory with three quark flavours and a global $SU(3)_L
\times SU(3)_R$ chiral symmetry 
\cite{Erlich:2005qh,DaRold:2005zs,Hirn:2005nr}.  We
work in the chiral limit and include the effect of the
spontaneous $\chi$SB via the IR boundary condition on the
axial vector fields \cite{DaRold:2005zs,Hirn:2005nr}. This
is clearly a rather severe limit, yet the calculational
simplification it affords makes it worthwhile. (The effects of 
spontaneous and explicit $\chi$SB
could be treated more precisely by the inclusion of a bulk
scalar in the bi-fundamental representation of the gauge group
\cite{Erlich:2005qh,DaRold:2005zs}.)

\subsection{Propagators and 5D Interactions}

By the holographic correspondence, the UV boundary values of $V_\mu$
and $A_\mu$ act as classical sources coupled to the 4D vector and
axial global symmetry currents.
Moreover, the 5D action evaluated on the solutions of
the equations of motion (EOM) of the bulk fields defines the
generating functional for current-current correlators in the 4D theory. 
We therefore need to solve the EOM with given UV
boundary values of the fields, and substitute back into the
action. The EOM are non-linear, so we must solve them iteratively.
This is the origin of Witten diagram construction for
calculating the $n$-point Greens functions of operators in the dual
theory \cite{Gubser:1998bc}. The iterative solution of the EOM requires the
bulk-to-bulk and bulk-to-boundary propagators for the gauge fields
in the 5D theory. 

We work in the unitary gauge in both the $SU(3)_V$ and $SU(3)_A$
sectors. $V_5$ is eliminated completely in this gauge if we impose
null Dirichlet boundary conditions for it on both branes. We impose warped
Neumann boundary conditions on the $A_5$ field (i.e. $\partial_{z}(aA_5)=0$
on both branes). In this gauge $A_5$ is a physical
field because it has a zero mode which
cannot be gauged away. The
$V_\mu$ and $A_\mu$ propagators have both transverse and
longitudinal parts in this gauge. Following Ref.\cite{Randall:2001gb} we have
\begin{equation}
\langle V^\mu V^\nu \rangle= - i G^V_p(z,z') P_T^{\mu\nu} 
                             - i G^V_0(z,z') P_L^{\mu\nu}, 
\end{equation}
where $P_T^{\mu\nu}=\eta^{\mu\nu}-\frac{p^\mu p^\nu}{p^{2}}$ 
and $P_L^{\mu\nu}=\frac{p^\mu p^\nu}{p^{2}}$.
The $A_\mu$ propagator has a similar form, and both solve
\begin{equation}
\left(\partial_z^2-\frac{1}{z}\partial_z+p^2 \right)G^{V,A}_p(z,z')=
\frac{z \, \delta(z-z')}{M_5L}. 
\end{equation}
The boundary conditions on $G^V_p(z,z')$ are Dirichlet on the UV brane, 
$G^V_p(z,z')\big\vert_{z=L_0}=0$, and Neumann on the IR brane, 
$\partial_z G^V_p(z,z')\big\vert_{z=L_1}=0$. 
We impose Dirichlet boundary conditions on both branes in the axial
sector $G^A_p(z,z')\big\vert_{z=L_0, L_1}=0$, to encode $\chi$SB. The
bulk-to-boundary propagator is defined as:
\begin{equation}
\langle V^\mu V^\nu \rangle\bigg\vert_{\partial ADS}(z')=-\frac{M_5L}{z}\partial_z\langle V^\mu
V^\nu\rangle\Big\vert_{z=L_0}, 
\end{equation}
(with a similar equation for $A_\mu$), where
\begin{equation}
\langle V^\mu V^\nu \rangle\bigg\vert_{\partial ADS}(z')= -iK^V_p(z') P_T^{\mu\nu}
                                                          -iK^V_0(z') P_L^{\mu\nu}, 
\end{equation}
\begin{eqnarray}
K^{V,A}_p(z') & = & -\frac{M_5L}{z}\partial_z G^{V,A}_p(z,z')_{z<z'}\Big\vert_{z=L_0}  \, , 
\end{eqnarray}
and similarly for $K^{V,A}_0(z')$. The solutions are given by:
\begin{eqnarray}
&&G^{V,A}_p(z,z')_{z<z'}= \pi z z' \times \nonumber \\
&&\frac{[A\mathcal{J}_1(pz)+B\mathcal{Y}_1(pz)]
[C\mathcal{J}_1(pz')+D\mathcal{Y}_1(pz')]}{2M_5L(AD-BC)},
\end{eqnarray}
$z>z'$ implies the exchange $z\leftrightarrow z'$. 
For the vector sector the coefficients are
$A=-\mathcal{Y}_1(pL_0)$, $C=-\mathcal{Y}_0(pL_1)$, 
$B=\mathcal{J}_1(pL_0)$ and $D=\mathcal{J}_0(pL_1)$;
for $A_\mu$ we obtain $A=-\mathcal{Y}_1(pL_0)$, $C=-\mathcal{Y}_1(pL_1)$ 
$B=\mathcal{J}_1(pL_0)$ and $D=\mathcal{J}_1(pL_1)$.
In the zero momentum limit:
\begin{equation}
G^V_0(z,z')_{z<z'}=-\frac{1}{2M_5L}(z^2-L_0^2), 
\end{equation}
($z>z'$ again swaps $z$ with $z'$). For the
axial sector
\begin{equation}
G^A_0(z,z')_{z<z'}=-\frac{1}{2M_5L}  \frac{(z^2-L_0^2) \, (z'^2-L_1^2)}
{L_0^2-L_1^2},
\end{equation}
and swapping $z$ and $z'$ for $z>z'$. The bulk-to-boundary propagators are then
\begin{equation}
K^{V,A}_p(z')  =  -\frac{z'}{L_0}\frac{C\mathcal{J}_1(pz')+D\mathcal{Y}_1(pz')}{AD-BC},
\end{equation}
while $K^V_0(z') =  1$, and $K^A_0(z') = \frac{z'^2-L_1^2}{L_0^2-L_1^2}$.
The physical condition of 4D current conservation eliminates the 
longitudinal piece
of the bulk-to-boundary propagators.
The positions of the poles of $K^V_P$ and $K^A_P$ are the masses
of the vector and axial-vector mesons, respectively. This is the
usual association of the masses of the Kaluza-Klein modes in 5D
with the masses of the bound states of the dual theory
\cite{Arkani-Hamed:2000ds}. Note that this tree-level treatment
of the 5D model incorrectly
predicts that the mass of the $n$'th excited resonance scales
like $n$ (for $n\gg 1$) as opposed to the $\sqrt{n}$ Regge behaviour.
This failing is to be expected, as the simple 5D theory here
presented is invalid in the far UV \cite{Karch:2006pv}.

We also note that the current-current correlator in the axial
sector as $p\to 0$ is
$\Pi_A(p^2)\vert_{p=0}= F_{\pi}^2 =2M_5L/(L_1^2-L_0^2)$
\cite{Erlich:2005qh,DaRold:2005zs,Hirn:2005nr}.
The $A_5$ propagator in this gauge is $\langle A_5 A_5 \rangle= -iG^5_p(z,z')$,
where $G^5_p(z,z')$ is the limit as $\xi \to \infty$ of the solution of
\begin{equation}
\left(\xi\partial_z^2-\xi\frac{1}{z}\partial_z+\xi\frac{1}{z^2}+p^2
\right)G^5_p(z,z')=-\frac{z \, \delta(z-z')}{M_5L} \, ,
\end{equation}
with the boundary conditions $\partial_z(aG^5_p(z,z'))\big\vert_{z=L_0, L_1}=0$,
giving
\begin{equation}
G^5_p(z,z')=\frac{-zz'}{M_5L (L_1^2-L_0^2)} \, \frac{1}{p^2} .
\end{equation}
The interactions are the three-gauge boson and four-gauge boson vertices
derived from the 5D action Eq.(\ref{action}).

\subsection{Four-point current-current correlators}

In order to calculate the four-point current-current correlators we have to
solve tree level four-point Witten diagrams which contain either
one four-gauge field vertex or two three-gauge field vertices.
Note that certain 5d Witten diagrams do not contribute to the four-point functions
(\ref{WLRLR}) and  (\ref{LLRRint}) because they do not preserve the flavor structure of
the 4d theory.
\begin{figure}
\includegraphics[width=7cm]{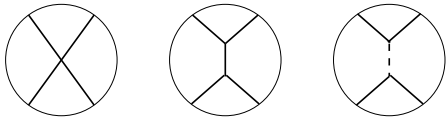}
\caption{\label{fig:wittendiags} The three basic structures of the 36 individual
Witten diagrams: X-diagrams involving the 4-boson vertex, Y-diagrams
involving two 3-boson vertices, and Y diagrams where $A_5$ propagates.  The
exterior circle represents the (UV)
boundary of the AdS$_5$ space, the solid lines are boundary-to-bulk
or bulk-to-bulk (axial or vector) gauge propagators, and the dotted line
is a bulk-to-bulk $A_5$ propagator.}
\end{figure}

We find that the group theory structure is a common factor
in all diagrams,
so denoting by $\Sigma$ the sum of all diagrams with group factors amputated
we have $\Sigma=\Sigma_{X}+\Sigma_{A_5}+\Sigma_{Y}$, with
\begin{eqnarray}
&&\! \Sigma_{X}=3iM_5L\int\frac{\mathrm{d}z}{z}
\bigl([{K_0^V}^2+{K_0^A}^2][{K_p^V}^2+{K_p^A}^2]~\nonumber\\
&&~~~~~~~~~~-4K_0^VK_0^AK_p^VK_p^A\bigr)\\
&&\! \Sigma_{A_5}=-\frac{3 i}{2}(M_5L)^2\int
\frac{\mathrm{d}z}{z}\frac{\mathrm{d}z'}{z'}G^5_p(z,z')A'(z,z')~
\end{eqnarray}
where $A'(z,z')$ is a function whose explicit form is known.
As for the Y-diagrams, the integrations are more involved, but the sum can be
written as $\Sigma_{Y}=Y_1+Y_2+Y_3+Y_4-2Y_5-2Y_6$, where 
\begin{eqnarray}
&&Y_i =-3i\left(\frac{M_5L}{\sqrt{2}}\right)^2 p^2\int\frac{\mathrm{d}z}{z}
\frac{\mathrm{d}z'}{z'}\nonumber \\
&&~~~~K_p^{a_i}(z)G_0^{b_i}(z,z')K_p^{c_i}(z')K_0^{d_i}(z)K_0^{e_i}(z') 
\end{eqnarray}
with $\{ abcde \} = \{ VVVVV \}, \{ VAVAA \}, \{ AAAVV \}$,
$\{ AVAAA \}, \{ AAVVA \}, \{ AVVAV \}$ for $i=1,...,6$. 
Performing the Wick rotation to Euclidean space, we obtain:
\begin{equation}
W_{LRLR}(Q^2) = \frac{i Q^2}{6 F_\pi^2}\Sigma(p=iQ) =-2 W_{LLRR}(Q^2) .
\label{Wresult}
\end{equation}
Thus we find that the ratio of the two correlators is independent of the
integration in $z$-space, and is solely due to their different 
$SU(3)_L \times SU(3)_R$ symmetry factors. 
The integrals can be done analytically with finite values for $L_1$ and $L_0$,
but our observables have a smooth $L_0\rightarrow 0$ limit, and much simpler
expressions arise if we set $L_0=0$ giving:
\begin{eqnarray}
\Sigma(Q)&=&3iM_5L\left[\frac{32}{Q^6L_1^6}-\frac{16}{5Q^4L_1^4}-\frac{69}{60I_1^2}+\frac{2}{5Q^2L_1^2I_1^2} \right. \nonumber \\
{} &+& \frac{69}{60I_0^2}-\frac{3}{15Q^2L_1^2I_0^2}+\frac{8}{5Q^4L_1^4I_0^2} \nonumber \\
{} &+& \left.\frac{32}{Q^6L_1^6I_0^2}-\frac{64}{Q^6L_1^6I_0}+\frac{1}{QL_1I_0I_1}\right] , 
\label{siresult}
\end{eqnarray}
where $I_{0,1}\equiv I_{0,1}(QL_1)$ are modified
Bessel functions of zeroth and first order respectively.

\section{Results and Discussion} 

The three input parameters of our model are $M_5L$, $L_0$ and $L_1$.  From the expressions
in Section II.A one can easily compute the boundary--to--boundary propagator in the
vector sector, 
and match its large Euclidean momentum behaviour to that known
from perturbative QCD deriving the relation\cite{Erlich:2005qh, DaRold:2005zs}
\begin{equation}
M_5L = \frac{N_c}{12 \pi^2} .
\label{Ncreln}
\end{equation}
One is then left with the two dimensionful parameters, $1/L_0$ giving the UV
cutoff scale, and $1/L_1$ setting the IR scale.  As mentioned before,
one problem of the simple 5D holographic model presented is that it does not
correctly reproduce the spectrum of meson resonances in the regime above approximately 1500 MeV. 
In addition, in computing the physically relevant observables $g_8$ and $g_{27}$ one
must in practice choose a scale $\mu$ at which to evaluate the Wilson coefficients $z_1(\mu)$
and $z_2(\mu)$.  One simple choice we make in the following is 
to take $1/L_0=1500$~MeV, imposing this as an upper hard cutoff when we evaluate the $Q^2$ integrals 
in Eqs.(\ref{BKint})-(\ref{LLRRint}), 
and identifying this hard cut-off with the short distance 
renormalisation scale $\mu$.
Although this matching procedure is quite 
crude, Refs.~\cite{BBGandco} have shown
that such a procedure captures the dominant contributions.  In a later 
longer publication
we investigate in some detail other more sophisticated matching procedures.
With respect to the $\chi$PT and perturbative QCD calculations of \cite{BBGandco} and
\cite{PR,Hambye:2003cy} of the $B_K$ and $\Delta I=1/2$ rule observables, our model,
if valid, presents the interesting possibility of calculating the intermediate $Q^2$ 
contribution around 0.1-2GeV$^2$ directly, rather than fitting it from an 
interpolation between the $\chi$PT low $Q^2$ regime \cite{BBGandco} and of 
the OPE high $Q^2$ regime \cite{PR,Hambye:2003cy}.

One immediate check on our calculation follows from the fact that the most general form
for the Greens function $W_{LRLR}(Q^2)$ in the large $N_c$ limit
can be written in terms of the masses, $M_i$, and residues
of the meson resonances \cite{PR,Hambye:2003cy}
\begin{equation}
\sum_{i=1}^{\infty}\left(\frac{\alpha_i}{(Q^2+M^2_{i})}+\frac{\beta_i}
{(Q^2+M^2_{i})^2}+\frac{\gamma_i}{(Q^2+M^2_{i})^3}\right),\label{eq6}
\end{equation}
and similarly for $W_{LLRR}$ with residues $\alpha'_i$, $\beta'_i$
and $\gamma'_i$.  We have explicitly checked that our results for 
$W_{LRLR}(Q^2)$ and $W_{LLRR}(Q^2)$ agree with this general form.

We now turn to the comparison of our results with the experimental
data.  Given Eq.(\ref{Ncreln}) and our choice of $L_0=1/1500$ MeV$^{-1}$,
the one remaining free parameter of our model is $L_1$.  The data to which this
must be fitted are, most importantly, the pion decay
constant, $F_\pi$, the $\rho$ vector-meson mass, $m_\rho$, the $a_1$ axial-vector meson
mass $m_{a_1}$, and the $g_8$ and $g_{27}$ (or $B_K$) parameters.   In the low-$Q^2$
regime chiral perturbation theory calculations
\cite{PR,Hambye:2003cy} give the behaviour of the 
correlation functions
\begin{eqnarray}
W_{LRLR}(Q^2)&=&6-24({2 l_1+ 5 l_2 + l_3 +l_9})\frac{Q^2}{F_\pi^2}+...
\nonumber\\
W_{LLRR}(Q^2)&=&-\frac{3}{8}
+(-\frac{15}{2}l_{3}+\frac{3}{2}l_9)\frac{Q^2}{F_{\pi}^2}+...
\label{chptresults}
\end{eqnarray}
with $l_i$ the standard chiral-Lagrangian coefficients.   In our fit procedure we use
the $\chi$PT results, Eq.(\ref{chptresults}), in the integrals over $Q^2$ below $1/L_1$, and the results
of the holographic calculation, Eqs.(\ref{Wresult}) and (\ref{siresult}) in the regime $1/L_1$
to $1/L_0$.

In terms of $M_5L$, $L_1$ and $L_0$ the predictions of the 5D holographic model for $F_\pi$ and
$m_\rho$ and $m_{a_1}$ are \cite{Erlich:2005qh,DaRold:2005zs,Hirn:2005nr}
\begin{equation}
F_\pi^2 \approx \frac{2M_5 L}{L_1^2- L_0^2}
\label{fpi}
\end{equation}
and in a good approximation in the range of interest
\begin{eqnarray}
m_\rho &\approx & \frac{2.12}{L_1} \frac{(L_1 -0.282 L_0)}{(L_1-L_0)} \\
m_{a_1} & \approx & \frac{3.38}{L_1} \frac{(L_1 -0.085 L_0)}{(L_1-L_0)} .
\label{mrho_ma1}
\end{eqnarray}
(We use a precise interpolation to the exact result for our fits.)
A non-trivial success of the 5D holographic QCD model is that in the $L_0=0$ limit (to which
$1/L_0 =1500$ MeV is a good approximation) the ratio $m_{a_1}/m_\rho \approx 1.6$ is very
well reproduced independently of the values of $M_5L$ and $L_1$, as already noted in 
\cite{Erlich:2005qh,DaRold:2005zs}.  

Numerically performing the $Q^2$ integrations in Eqs.(\ref{BKint}), (\ref{int2}), and (\ref{LLRRint}), 
as a function of $1/L_1$ and utilizing, as appropriate, the leading order values of the
Wilson coefficients $z_1$ and $z_2$ (taken from \cite{Buras} 
with $\Lambda_{\bar{MS}}^{(4)}=325$~MeV, and $\mu =1500$~MeV) we
can fit the final parameter $L_1$ to the full set
of observables.  The $\Delta I = 1/2,3/2$ data 
requires $g_8|_{\underline{obs}}=5.1$ 
and $g_{27}|_{\underline{obs}}=0.29$, although the values 
to be explained are modified to $g_8|_{obs}=3.3$ and $g_{27}|_{obs}=0.23$
when one takes into account 
the enhancement already provided by 
the calculated ${\cal{O}}(p^4)$ chiral corrections \cite{KMW}. 
Fitting to $g_8$ and $g_{27}$ and the values of $F_\pi$, $m_\rho$ and
$m_{a_1}$ leads to the result
\begin{equation}
L_1^{-1} \approx 302~{\rm MeV}
\label{params}
\end{equation}
which implies (in all cases normalized to the data for ease of understanding)
\begin{eqnarray}
&&\frac{m_\rho |_{th}}{m_\rho |_{obs}} \approx 1.00,\,\,\, 
\frac{m_{a_1} |_{th}}{m_{a_1} |_{obs}} \approx 1.04,\,\,\,
\frac{F_\pi |_{th}}{F_\pi |_{obs}} \approx 0.80, \nonumber \\
&&\frac{g_8 |_{th}}{g_8 |_{obs}}\approx 0.49, \,\,\,
\frac{g_{27} |_{th}}{g_{27} |_{obs}} \approx 1.39
\label{fitresults}
\end{eqnarray}
and a value of $B_K(1500\,\hbox{MeV}) \approx 0.54$ ($\hat{B}_K \approx0.76$
using the Wilson coefficient 
value $C_{\Delta S = 2}(1500$ MeV) $= 1.42$, see e.g.~Ref.~\cite{PR}). 
Considering the relative crudity of the model and the use
of the large-$N_c$ expansion of QCD 
this is a reasonable fit
to the data and we find this result very encouraging. We emphasise that $L_1$ and $L_0$ are 
the only free parameters in this fit to the five observables of equation Eq.(\ref{fitresults}).  
It is noteworthy that in the 
expected domain of validity of the 
model (i.e. in the $Q^2$ intermediate region), the 5D interactions
induce a large increase of $g_8$ and a suppression for $g_{27}$.

On the other hand
the holographic model predicts the low-$Q^2$
behaviour of $W_{LRLR}$ and $W_{LLRR}$ to be identical in form to $\chi$PT
but with disagreeing numerical values 
\begin{equation}
W_{LLRR}=-\frac{W_{LRLR}}{2}=-\frac{1}{4}+\frac{M_5L}{4}
\frac{29}{24}\frac{Q^2}{F_\pi^2}+...
\end{equation}
It is possible
that this failing is due to the simplistic truncation of the AdS space
in the IR. Alternatively, the fault might be with our
approximations, particularly the treatment in the 5D model of $\chi$SB.


It would be interesting to include a bi-fundamental bulk scalar
field of mass-squared $-3/L^2$ associated with the ${\bar q} q$
operator, as in the models \cite{Erlich:2005qh,DaRold:2005zs}, and
a massless singlet scalar associated with the glueball states
generated by the gluon field
operator $G_{\mu\nu}G^{\mu\nu}$, to see whether their inclusion 
leads to an even better fit to the full set of observables.

\begin{acknowledgments}
This work has been supported by PPARC Grant PP/D00036X/1, and
by the `Quest for Unification' network, MRTN 2004-503369. 
The work of BH was supported by the Clarendon Fund and the one of TH by 
a Ramon y Cajal contract of the Ministerio de 
Educaci\'on y Ciencia.

\end{acknowledgments}

\end{document}